\documentclass[10pt]{article}
\begin{document}

%%%%%%%%%%%%%%%%%%%%
\title{{\bf{\Large  Non(anti) commutativity for open superstrings}}}
%%%%%%%%%%%%%%%%%%%%
\author{
{\bf {\normalsize Biswajit Chakraborty}$^{a,b,}$\thanks{biswajit@bose.res.in}},\,
 {\bf {\normalsize Sunandan Gan{g}opadhyay}$^{a,}$\thanks{sunandan@bose.res.in}} \\
 {\bf {\normalsize Arindam Ghosh Hazra}$^{a,}$\thanks{arindamg@bose.res.in}},\,\,{\bf {\normalsize Frederik G. Scholtz}$^{a,b,}$\thanks{fgs@sun.ac.za}}
\\
$^{a}$ {\normalsize S.~N.~Bose National Centre for Basic Sciences,}\\{\normalsize JD Block, Sector III, Salt Lake, Kolkata-700098, India}\\[0.3cm]
$^{b}$ {\normalsize Institute of Theoretical Physics, University of
Stellenbosch,}\\{\normalsize Stellenbosch 7600, South Africa}\\[0.3cm]
}
\date{}

\maketitle

%%%%%%%%%%%%%%%%
\begin{abstract}
Non(anti)commutativity in an open free superstring and also one 
moving in a background anti-symmetric tensor field is investigated.
In both cases, the non(anti)commutativity is shown to be a direct
consequence of the non-trivial boundary conditions which, contrary
to several approaches, are not treated as constraints. The above
non(anti)commutative structures lead to new results in the
algebra of super constraints which still remain involutive,
indicating the internal consistency of our analysis.
%%%%%%%%%%%%%%%%

\\[0.3cm]

\end{abstract}

%---------------------------------------
\section{Introduction}
In the last few years, there has been a considerable interest in 
the study of open strings propagating in the presence
 of a background Neveu-Schwarz two-form
field $B_{\mu\nu}$, leading to a noncommutative  structure \cite{sw,dn}.
 This structure manifests in the
noncommutativity in the spacetime coordinates of D-branes, 
where the end points of the open strings
are attached. Different approaches have been
adopted to obtain this result in the case of both the bosonic as well as the
fermionic superstring. A Hamiltonian operator treatment was
provided in \cite{chu} and a world sheet approach in \cite{som}. These
studies have been done in the bosonic theory. An
alternative Hamiltonian (Dirac \cite{di}) approach based on
 regarding the Boundary Conditions (BC)
as constraints was given in \cite{ard, bra}, investigations being carried
 out in both the bosonic and fermionic string theories.
%; the corresponding Lagrangian (symplectic)
%version being done in \cite{bra}.
 The interpretation of the BC as
primary constraints usually lead to an infinite tower
 of second class constraints
\cite{tez}, in contrast to the usual Dirac formulation of constrained
systems \cite{di, hrt}. Besides, in this approach, where one tries 
to obtain non-commutativity through Dirac brackets between coordinates,
one encounters ambiguous factor like $\delta(0)$. Furthermore,
different results are obtained depending on the interpretations
of these factors \cite{ard}.

On the other hand, it has also been shown, by one of the authors,
that non-commutativity can be obtained in a more transparent manner
by modifying the cannonical Poisson bracket structure, so that it is 
compatible with the boundary condition \cite{rb}. In this approach,
the boundary conditions are {\it{not}} treated as constraints. 
This is similar in spirit to the treatment of Hanson, Regge and 
Teitelboim \cite{hrt}, where modified PBs were
obtained for the free NG string, in the orthonormal gauge, which is
 the counterpart of the conformal gauge in the free Polyakov string.
Those studies were, however, restricted to the case of the bosonic string
and membrane only. We extend the same methodology to the superstring
in this paper.

Some other approaches to this problem have also been 
discussed in \cite{zab, and}. As has been stressed in \cite{sw}, it is very important to
understand this noncommutativity from different perspectives.

%--------------------------------------------------------
We find that the super-Virasoro constraints play a crucial
role in revealing the non (anti) commutative
structure. 
The paper is organized as follows: In section {\bf 2}, the RNS
superstring action in the conformal gauge
 is discussed. This also helps to fix the notations.
 In section {\bf 3}, the boundary conditions of the fermionic sector
of the superstring is given and the non-anticommutativity
 of the theory
 is revealed in the
conventional hamiltonian framework. The results are also tied up with
 the bosonic theory.  
  Section {\bf 4} discusses the non (anti) commutativity in the interacting
superstring theory in the RNS formulation. The paper ends with a
 conclusion in
section {\bf 5}.

%--------------------------------------- 

%%%%%%%%%%%%%%%%%%%%%%%%%%%%%%%%%%%%%%%%%%%%%%%%%%%%%%%%%%%%%%%%%%%%%%%%%%%%
%%%%%%%%%%%%%% abstract
%%%%%%%%%%%%%%%%%%%%%%%%%%%%%%%%%%%%%%%%%%%%%%%%%%%%%%%%%%%%%%%%%%%%

\section {Free superstring}

Let us consider the action for the free superstring, in conformal 
gauge \cite{green},

\begin{eqnarray}
\label{4}
S \,=\,
{i \over 4 }
\int_{\Sigma} d^2\sigma \,d^2 \theta \Big(
{\overline D} Y^\mu D Y_\mu  \Big)\,,
\end{eqnarray}

\noindent where the superfield 
 \begin{eqnarray}
Y^\mu (\sigma, \theta) \,=\,
 X^\mu (\sigma) 
+ {\overline \theta}  \psi^\mu (\sigma) + \frac{1}{2} 
{\overline \theta} \theta B^\mu  (\sigma)
\label{4ka}
\end{eqnarray}
\noindent unites the bosonic ($X{^\mu}(\sigma)$) and fermionic
 ($\psi^{\mu}(\sigma)$) spacetime string coordinates with 
a new auxiliary bosonic field $B{^\mu}(\sigma)$.

In component form the action reads \footnote{Our conventions are: 
$\rho^0 \,=\, \sigma^2 \, = \, 
\pmatrix{0&-i\cr i&0\cr}\,\,\,,\,\,\,
\rho^1 \,=\, i\sigma^1 \, = \, \pmatrix{0&i\cr i&0\cr}\ $.
Our signature of the 
induced world-sheet metric and target space-time metric are 
$\eta^{ab} = \{-, +\}$,  
 $\eta^{\mu\nu} =\{-, +, +, ...., +\}$ respectively and
  $\bar{\theta}$ is defined as
$\bar{\theta} = \theta^T\rho^0$.}

\begin{eqnarray}
\label{1}
S &=& -\frac{1}{2}
\int_{\Sigma} d^2\sigma \Big(
\eta_{\mu \nu } \partial_a X^\mu \partial^a X^\nu 
\,- \,  i {\overline \psi}^\mu \rho^a \partial_ a \psi_\mu \Big)\\
 &=& S_B + S_F \,\nonumber
\end{eqnarray}
where
\begin{eqnarray}
\label{1aa}
 S_B = -\frac{1}{2}\int_{\Sigma} d^2\sigma \eta_{\mu \nu }
 \partial_a X^\mu \partial^a X^\nu\, ,\quad 
S_F = \frac{1}{2}\int_{\Sigma} d^2\sigma
 i {\overline \psi}^\mu \rho^a \partial_ a \psi_\mu  
\end{eqnarray}
represent the decoupled bosonic and fermionic actions,
respectively. The fermions are taken to be  
Majorana and we refer to the component of $\psi$  as $\psi_{\pm}$
(compatible with our conventions)
\begin{equation}
\psi^\mu \,=\, \pmatrix{\psi^\mu_{-}\cr \psi^\mu_{+}\cr}\, .
\label{3}
\end{equation}

\noindent The equal time canonical antibrackets read,
 in terms of the components of $\psi$,
\begin{eqnarray}
\{ \psi^\mu_{+} (\sigma) \,,\,\psi^\nu_{+}  (\sigma^\prime) \}_{D.B} &=& 
\{ \psi^\mu_{-} (\sigma) \,,\,\psi^\nu_{-} (\sigma^\prime ) \}_{D.B}
\,\,=\,\,
 - i \eta^{\mu\nu} \delta (\sigma - \sigma^\prime )\,,
\nonumber\\
\{ \psi^\mu_{+}(\sigma)\,,\,\psi^\nu_{-}(\sigma^\prime) \}_{D.B} &=& 0\,\,.
\label{5}
\end{eqnarray}
This, along with the brackets 
\begin{eqnarray}
\label{18a}
\{X^\mu(\sigma) , \Pi^\nu(\sigma^{\prime})\} = 
 \eta^{\mu \nu}\delta(\sigma - \sigma^{\prime})
\end{eqnarray}
from the bosonic sector, defines the preliminary symplectic 
structure of the theory ($\Pi^\mu$ is the cannonically conjugate momentum
to $X^\mu$, defined in the usual way).

Confining our attention to $S_F$ (\ref{1aa}),
we vary the action (\ref{1aa})
\begin{eqnarray}
\label{18}
\delta S_F = i \int_{\Sigma} d^2\sigma \left[ \rho^a \, \partial_a 
\psi^{\mu} \, \delta \bar{\psi}_{\mu} - \partial_{\sigma}
\left(\psi^\mu_{-}\, \delta \psi_{\mu -} - 
\psi^\mu_{+}\, \delta \psi_{\mu +}\right)\right]
\end{eqnarray}
to obtain the Euler-Lagrange equation for the fermionic field 
\begin{eqnarray}
\label{10}
i \rho^a \partial_a \psi^{\mu} =0.
\end{eqnarray}

%------------------------------------------------------
The total divergence term yields the necessary BC. We shall consider 
its consequences in the following sections where the 
preliminary (anti) brackets will be modified.
Using the standard Noether procedure \footnote{We now use the supersymmetry 
transformations on-shell and hence we drop the auxilliary field
$B^{\mu}$ henceforth.}, the forms
of the supercurrent and the energy-momentum tensor (which are
constraints themselves \cite{green}) can be derived.
The expressions are:
\begin{eqnarray}
J_{a}=-\frac{1}{2}\rho^{b}\rho_{a}\psi^{\mu}\partial_{b}X_{\mu} = 0\,, 
\label{17a}
\end{eqnarray}
\begin{eqnarray}
T_{ab}=\partial_{a}X^{\mu}\partial_{b}X_{\mu}
-\frac{i}{4}\bar{\psi^{\mu}}\rho_{a}\partial_{b}\psi_{\mu}
-\frac{i}{4}\bar{\psi^{\mu}}\rho_{b}\partial_{a}\psi_{\mu}
-\frac{1}{2}\eta_{ab}(\partial^{c}X^{\mu}\partial_{c}X_{\mu}+
\frac{i}{2}\bar{\psi^{\mu}}\rho^{a}\partial_{a}\psi_{\mu}) = 0\,.
\label{17b}
\end{eqnarray}
All the components of $T_{ab}$ are, however, not independent as the energy-
momentum tensor is traceless
\begin{eqnarray}
T^{a}_{\ a} = \eta^{ab}T_{ab} = 0\,,
\label{17c}
\end{eqnarray}
leaving us with only two independent components of $T_{ab}$.
These components, which are the constraints of the theory,
are given by
\begin{eqnarray}
\chi_1(\sigma) = 2T_{00} = 2T_{11} &=& \Phi_1(\sigma) + \lambda_1(\sigma)
= 0, \nonumber \\ 
\chi_2(\sigma) = T_{0 1} &=& 
\Phi_2(\sigma) + \lambda_2(\sigma) = 0,
\label{27}
\end{eqnarray}
where
\begin{eqnarray}
\Phi_1(\sigma) &=& 
\left(\Pi^2(\sigma) + (\partial_{\sigma}X(\sigma))^2\right), \nonumber \\ 
\Phi_2(\sigma) &=& 
\left(\Pi(\sigma)\partial_{\sigma}X(\sigma)\right), \nonumber \\ 
\lambda_1(\sigma)&=&  
- i \bar{\psi^{\mu}}(\sigma)\rho_{1} \partial_{\sigma} \psi_{\mu}(\sigma)
=  - i \left(\psi^{\mu}_{-}(\sigma) 
\partial_{\sigma} \psi_{\mu -}(\sigma) - \psi^{\mu}_{+}(\sigma) 
\partial_{\sigma} \psi_{\mu +}(\sigma)\right), \nonumber \\ 
\lambda_2(\sigma)&=& 
- \frac{i}{2}  \bar{\psi^{\mu}}(\sigma)\rho_{0} 
\partial_{\sigma} \psi_{\mu}(\sigma)  
= \frac{i}{2}\left(\psi^{\mu}_{-}(\sigma)\partial_{\sigma} \psi_{\mu -}(\sigma) + \psi^{\mu}_{+}(\sigma)\partial_{\sigma} \psi_{\mu +}(\sigma)\right).
\label{27q}
\end{eqnarray}
The role of these constraints in generating those infinitesimal
diffeomorphisms which do not lead out of the conformal gauge is
well known \cite{green} and we are not going to elaborate on this. 
Note that the constraints that we obtain in this paper are on-shell,
i.e. we have used the equation of motion (\ref{10})
 for the fermionic field $\psi$.
This allows us to write them down in terms of the phase-space variables
\footnote{This is in the true spirit of Dirac's classic analysis of
constrained hamiltonian dynamics \cite{di}.}
and hence they look quite different from the standard results found in
the literature \cite{green}where they are written down in 
the light-cone coordinates which involves time derivatives.

From the basic brackets (\ref{5}), it is easy to generate a closed
(involutive) algebra:
\begin{eqnarray}
\{\chi_1(\sigma) , \chi_1(\sigma^{\prime})\} &=& 4 \left(
\chi_2(\sigma) + \chi_2(\sigma^{\prime})\right)\partial_{\sigma}
\delta(\sigma - \sigma^{\prime})\,,\nonumber \\ 
\{\chi_2(\sigma) , \chi_2(\sigma^{\prime})\} &=& \left(
\chi_2(\sigma) + \chi_2(\sigma^{\prime})\right)\partial_{\sigma}
\delta(\sigma - \sigma^{\prime})\,,\nonumber\\
\{\chi_2(\sigma) , \chi_1(\sigma^{\prime})\} &=& \left(
\chi_1(\sigma) + \chi_1(\sigma^{\prime})\right)\partial_{\sigma}
\delta(\sigma - \sigma^{\prime})\,.
\label{4200}
\end{eqnarray}
It is interesting to observe that the structure of the super 
constraint algebra is exactly similar to the Bosonic theory \cite{rb}.

Coming to the super current $J_{aA}$
\footnote{$A = 1,2$ being the spinor index},
 note that it is a two component spinor. Further, since $J_{a}$ 
obeys the relation $\rho^a J_a = 0$, the components of $J_{0A}$ and
 $J_{1A}$ are related to each other. Hence we only deal with the
components of $J_{0A}$ or simply $J_1$ and $ J_2$. 
These are \footnote{$J_1$, $ J_2$ along with $\chi_1(\sigma)$ and 
$\chi_2(\sigma)$ constitutes the full set of super-Virasoro constraints}:
\begin{eqnarray}
\tilde{J}_{1}(\sigma) &=& 2J_{1}(\sigma) = 
(\psi^{\mu}_{-}(\sigma)\Pi_{\mu}(\sigma)
-\psi^{\mu}_{-}(\sigma)\partial_{\sigma}X_{\mu}) = 0\,, \nonumber\\
\tilde{J}_{2}(\sigma) &=& 2J_{2}(\sigma) = 
(\psi^{\mu}_{+}(\sigma)\Pi_{\mu}(\sigma)
+\psi^{\mu}_{+}(\sigma)\partial_{\sigma}X_{\mu}) = 0\,.
\label{280}
\end{eqnarray}
The algebra between the above constraints read:
\begin{eqnarray}
\{\tilde{J}_{1}(\sigma) , \tilde{J}_{1}(\sigma^{\prime})\}
&=&-i(\chi_{1}(\sigma)-2\chi_{2}(\sigma))
\delta(\sigma-\sigma^{\prime})\,,\nonumber\\
\{\tilde{J}_{2}(\sigma) , \tilde{J}_{2}(\sigma^{\prime})\}
&=&-i(\chi_{1}(\sigma)+2\chi_{2}(\sigma))
\delta(\sigma-\sigma^{\prime})\,,\nonumber\\
\{\tilde{J}_{1}(\sigma) , \tilde{J}_{2}(\sigma^{\prime})\}&=&0\,.
\label{281}
\end{eqnarray}
The algebra between $\tilde J(\sigma)$ and $\chi(\sigma)$ is also given by
\begin{eqnarray}
\{\chi_{1}(\sigma) , \tilde{J}_{1}(\sigma^{\prime})\}
&=& - \left(2 \tilde{J}_{1}(\sigma) + \tilde{J}_{1}(\sigma^{\prime})\right)
\partial_{\sigma}\delta(\sigma-\sigma^{\prime})\,,\nonumber\\
\{\chi_{1}(\sigma) , \tilde{J}_{2}(\sigma^{\prime})\}
&=& \left(2 \tilde{J}_{2}(\sigma) + \tilde{J}_{2}(\sigma^{\prime})\right)
\partial_{\sigma}\delta(\sigma-\sigma^{\prime})\,,\nonumber\\
\{\chi_{2}(\sigma) , \tilde{J}_{1}(\sigma^{\prime})\}
&=& \left(\tilde{J}_{1}(\sigma) + \frac{1}{2}\tilde{J}_{1}(\sigma^{\prime})\right)\partial_{\sigma}\delta(\sigma-\sigma^{\prime})\,,\nonumber\\
\{\chi_{2}(\sigma) , \tilde{J}_{2}(\sigma^{\prime})\}
&=& \left(\tilde{J}_{2}(\sigma) + \frac{1}{2}\tilde{J}_{2}(\sigma^{\prime})\right)\partial_{\sigma}\delta(\sigma-\sigma^{\prime})\,.
\label{282}
\end{eqnarray}

%%%%%%%%%%%%%%%%%%%%%%%%%%%%%%%%%%%%%%%%%%%%%%%%%%%%%%%%%%%%%%%%
%%%%%        Boundary conditions
%%%%%%%%%%%%%%%%%%%%%%%%%%%%%%%%%%%%%%%%%%%%%%%%%%%%%%%%%%%%%%%%%%
\section {Boundary conditions, super-Virasoro algebra and non(anti)
commutativity}
As in the case of bosonic variables \cite{rb}, Fermionic coordinates also require careful consideration of the surface terms arising in the variation of the action (\ref{18}) 
\footnote{A detailed treatment of the boundary conditions is given
in \cite{green}}. Vanishing of these surface terms requires that 
($\psi_{+}\delta\psi_{+} - \psi_{-}\delta\psi_{-}$) should 
vanish at each end point of the open string. This is satisfied by making
$ \psi_{+} = \pm \psi_{-}$ at each end. Without loss of generality
we set
\begin{eqnarray}
\label{12}
\psi^{\mu}_{+}(0 , \tau) = \psi^{\mu}_{-}(0 , \tau).
\end{eqnarray}
The relative sign at the other end now becomes meaningful
and there are two cases to be considered. In the first case
(Ramond(R) boundary conditions)
\begin{eqnarray}
\label{6}
\psi^{\mu}_{+}(\pi , \tau) = \psi^{\mu}_{-}(\pi , \tau)
\end{eqnarray}
and in the second case (Neveu-Schwarz (NS) boundary conditions)
\begin{eqnarray}
\label{13}
\psi^{\mu}_{+}(\pi , \tau) = - \psi^{\mu}_{-}(\pi , \tau).
\end{eqnarray}
Here we will work with Ramond boundary conditions.
Combining (\ref{12}) and (\ref{6}) we can write
\begin{eqnarray}
 \Big(\psi^\mu_{+} (\tau , \sigma)  - \psi^\mu_{-}
 (\tau , \sigma)\Big)\mid_{\sigma =0,\pi} = 0\,.
\label{7}
\end{eqnarray}
\noindent The mode expansion of the components of Majorana fermion
takes the form \cite{green}
\begin{eqnarray}
\psi^{\mu}_{-}(\sigma , \tau) &=& \frac{1}{\sqrt{2}}\sum_{n\in Z} d^{\mu}_{n}
e^{-in(\tau - \sigma )}\,, \nonumber \\ 
\psi^{\mu}_{+}(\sigma , \tau) &=& \frac{1}{\sqrt{2}}\sum_{n\in Z} d^{\mu}_{n}
e^{-in(\tau + \sigma )}\,.
\label{31}
\end{eqnarray}
The above mode expansions immediately leads to
\begin{eqnarray}
\psi^{\mu}_{-}(-\sigma , \tau) &=& \psi^{\mu}_{+}(\sigma , \tau)\,,
\label{31ba}
\end{eqnarray}
which further yields, using (\ref{6}),
\begin{eqnarray}
\psi^{\mu}_{\pm}(\sigma=-\pi , \tau)
=\psi^{\mu}_{\pm}(\sigma=\pi , \tau)
\label{30h}
\end{eqnarray}
in the R-sector
\footnote{In the NS sector, we obtain a 
anti-periodic boundary condition
$\psi^{\mu}_{-}(-\sigma , \tau)
=-\psi^{\mu}_{-}(\sigma , \tau)$ at $\sigma=\pi$.}.

In the bosonic sector, on the other hand, we have to enlarge the domain
of definition of the bosonic field $X^{\mu}$ as
\begin{eqnarray}
X^{\mu}(\tau , -\sigma) &=& X^{\mu}(\tau , \sigma) 
\label{37}
\end{eqnarray}
so that it is an even function and satisfies Neumann BC \cite{rb}. This
is in contrast to (\ref{31ba}).
Consistent with this, we must have
\begin{eqnarray}
\Pi^{\mu}(\tau , -\sigma) &=& \Pi^{\mu}(\tau , \sigma)\,,\nonumber \\ 
{X^{\mu}}^{\prime}(\tau , -\sigma) &=& -{X^{\mu}}^{\prime}(\tau , \sigma)\,.
\label{37aa}
\end{eqnarray}
Now, from  (\ref{31ba}, \ref{37}, \ref{37aa}), we note that
the constraints $\chi_1(\sigma) = 0$
and $\chi_2(\sigma) = 0$ are even and odd respectively under
$\sigma \rightarrow -\sigma$. This also enables us to increase the domain
of definition of the length of the string from ($0\leq \sigma \leq \pi $)
to ($- \pi \leq \sigma \leq \pi $).

We may then write the generator of all $\tau$ and $\sigma$ 
reparametrization as the functional \cite{hrt}
\begin{eqnarray}
L[f] = \frac{1}{2}\int_{0}^{\pi} d\sigma \{f_+(\sigma) \chi_1(\sigma)
+ 2 f_-(\sigma) \chi_2(\sigma)\}\,,
\label{32}
\end{eqnarray}
where $f_{\pm}(\sigma)=\frac{1}{2}(f(\sigma)\pm f(-\sigma))$ are by
construction even/odd function and  $f(\sigma)$ is an arbitary
differentiable function defined in the extended interval $[-\pi, \pi]$.
The above expression can be simplified to
\begin{eqnarray}
L[f] = \frac{1}{4}\int_{-\pi}^{\pi} d\sigma f(\sigma)[\{\Pi(\sigma)+
\partial_{\sigma}X(\sigma)\}^2+2i\psi^{\mu}_{+}\partial_{\sigma}
\psi_{\mu+}]
\label{33}.
\end{eqnarray}
 Coming to the generators $J_{1}$ and $J_{2}$, note that $J_{1}(-\sigma)
= J_{2}(\sigma)$ (\ref{280}). This enables us to write 
 down the functional $G[g]$ 
\begin{eqnarray}
G[g] &=& \int_{0}^{\pi} d\sigma(g(\sigma)J_{1}(\sigma)+
g(-\sigma)J_{2}(\sigma)) \nonumber \\
&=&\int_{-\pi}^{\pi} d\sigma g(\sigma)J_{1}(\sigma)
= \int_{-\pi}^{\pi} d\sigma g(-\sigma)J_{2}(\sigma)
\label{34}
\end{eqnarray}
for any differentiable function $g(\sigma)$, defined again in the extended interval $[-\pi,\pi]$.  These functionals (\ref{33}), (\ref{34}) generate the following super Virasoro algebra:
\begin{eqnarray}
\{L[f(\sigma)] , L[g(\sigma)]\} &=& L[f(\sigma)g^{\prime}(\sigma)
- f^{\prime}(\sigma)g(\sigma)]\,,\nonumber \\
\{G[g(\sigma)] , G[h(\sigma)]\} &=& -i L[g(-\sigma)h(-\sigma)]\,, \nonumber \\
\{L[f(\sigma)] , G[g(\sigma)]\} &=& G[f(\sigma)g^{\prime}(-\sigma)
-\frac{1}{2}f^{\prime}(\sigma)g(-\sigma)]\,.
\label{34a}
\end{eqnarray}
Defining 
\begin{eqnarray}
L_m = L[e^{-im\sigma}]\ \ \mathrm{and} \ \ G_n = G[e^{in\sigma}]\,,
\label{34y}
\end{eqnarray}
one can write down an equivalent form of the super-Virasoro algebra
\begin{eqnarray}
\{L_m , L_n\} &=& i (m-n) L_{m + n}\,,\nonumber \\
\{G_m , G_n\} &=& -i L_{m + n}\,,\nonumber \\
\{L_m , G_n\} &=& i \left(\frac{m}{2} - n\right)G_{m + n}\,.
\label{34b}
\end{eqnarray}
Note that we do not have a central extension here, as 
the analysis is entirely classical.

Coming back to the preliminary symplectic structure, given in (\ref{5}),
 we note that the boundary conditions (\ref{7}) are not 
compatible with  the brackets, although one could get the
super-Virasoro algebra (\ref{34a}) or (\ref{34b}) just by using
(\ref{5}) and (\ref{18a}).
 Hence the last of the brackets in 
(\ref{5}) should be altered suitably. A simple inspection 
suggests that 
\begin{eqnarray}
\{\psi^\mu_{+}(\sigma)\,,\,\psi^\nu_{-}(\sigma^\prime)\}
 = -i \eta^{\mu\nu} \delta (\sigma - \sigma^\prime )\,.
\label{8}
\end{eqnarray}
Although the bracket structures (\ref{5}) and (\ref{8}) agree with \cite{godinho} (in the free case), they 
can, however, not be regarded as the final ones. This is because the
 presence of the usual Dirac delta function $\delta(\sigma - \sigma^{\prime})$
implicitly implies that the finite physical range of 
$\sigma \in [0 , \pi]$ for the string has not been taken into account.
Besides, it is also not compatible with (\ref{31ba}). 
In \cite{rb}, the equal time commutators were
given in terms of certain combinations 
($\Delta_{+}(\sigma , \sigma^{\prime})$) of periodic delta
 function\footnote{The form of the
periodic delta function is given by $\delta_{P}(x-y) = \delta_{P}(x-y+ 2\pi)
=\frac{1}{2\pi} \sum_{n\in Z}e^{in(x-y)}$ and is related to the usual
Dirac $\delta$-function as $\delta_P(x-y) = 
\sum_{n\in Z}\delta(x - y + 2\pi n)$ \cite{schwinger}.}
\begin{eqnarray}
\{X^{\mu}(\tau , \sigma) ,  \Pi_{\nu}(\tau , \sigma^{\prime})
= \delta^{\mu}_{\nu} \Delta_{+}(\sigma, \sigma^{\prime})\,,
\label{00}
\end{eqnarray}
where
\begin{eqnarray}
\Delta_{\pm}\left(\sigma , \sigma^{\prime})
= \delta_{P}(\sigma - \sigma^{\prime})\right)
\pm\delta_{P}(\sigma + \sigma^{\prime})\,, 
\label{00z}
\end{eqnarray}
 rather than an
ordinary delta function to ensure compatibility with Neumann BC
 in the bosonic sector. Basically, there one has to identify
the appropriate `` delta function " for the physical range 
$[0 , \pi]$ of  $\sigma$ starting from the periodic delta function
$\delta_P(\sigma - \sigma^{\prime})$ for the extended (but finite)
range  $[-\pi , \pi]$ and make use of the even nature of the bosonic
variables $X^{\mu}$ (\ref{37}) in the extended interval.
 Furthermore, the occurence of $\delta_P(\sigma - \sigma^{\prime})$
itself was justified by the fact that a scalar field, subjected
to periodic BC in a one-dimensional box of length $2\pi$ has 
$\delta_P(\sigma - \sigma^{\prime})$, rather than the usual delta
function, occuring in the basic Poisson-bracket between the scalar 
field and its conjugate momentum $\Pi$.

We can essentially follow the same methodology here in the fermionic sector as $\psi^{\mu}_{\pm}(\tau , \sigma)$ also satisfy periodic
BC of period $2 \pi$ (\ref{30h}). The only difference with the 
bosonic case, apart from the Grassmanian nature of the latter,
 is that, instead of their even property (\ref{37}),
the components of Majorana fermions satisfy (\ref{31ba}). As
 we shall show now that this condition is quite adequate to identify
the appropriate delta-functions for the ``physical interval"
$[0 , \pi]$.

We start by noting that the usual properties of a delta function is also
satisfied by $\delta_{P}(x)$
\begin{eqnarray}
\int_{-\pi}^{\pi}dx^{\prime}\delta_{P}(x^{\prime}-x)f(x^{\prime})=f(x)
\label{40}
\end{eqnarray}
for any periodic function $f(x)=f(x+2\pi)$ defined in the interval
 $[-\pi, \pi]$. Hence one  can immediately write down the following
expressions for $\psi^{\mu}_{-}$ and $\psi^{\mu}_{+}$:
\begin{eqnarray}
\int_{0}^{\pi}d\sigma^{\prime}\left[\delta_{P}(\sigma^{\prime} +
\sigma)\psi^{\mu}_{+}(\sigma^{\prime}) + 
\delta_{P}(\sigma^{\prime}-\sigma)\psi^{\mu}_{-}(\sigma^{\prime})
\right] =\psi^{\mu}_{-}(\sigma)\,,
\label{41}
\end{eqnarray}
\begin{eqnarray}
\int_{0}^{\pi}d\sigma^{\prime}\left[\delta_{P}(\sigma^{\prime} 
+ \sigma)\psi^{\mu}_{-}(\sigma^{\prime}) + 
\delta_{P}(\sigma^{\prime}-\sigma)\psi^{\mu}_{+}(\sigma^{\prime})
\right] =\psi^{\mu}_{+}(\sigma)\,.
\label{42}
\end{eqnarray}
Combining the above equations and writing them in a matrix form, we get,
\begin{eqnarray}
\int_{0}^{\pi}d\sigma^{\prime}\Lambda_{AB}(\sigma, \sigma^{\prime}) 
\psi^{\mu}_{B}(\sigma^{\prime})=\psi^{\mu}_{A}(\sigma)\quad;\quad (A=-,+)\,,
\label{43}
\end{eqnarray}
where $\Lambda_{AB}(\sigma, \sigma^{\prime})$, defined by 
\begin{equation}
 \Lambda_{AB}(\sigma, \sigma^{\prime})\,=\, \pmatrix{\delta_{P}(\sigma^{\prime}-\sigma)&\delta_{P}(\sigma^{\prime}+\sigma)\cr
\delta_{P}(\sigma^{\prime}+\sigma)&\delta_{P}(\sigma^{\prime}-\sigma)\cr}\,,
\label{44}
\end{equation}
acts like a matrix valued ``delta function"
for the two component Majorana spinor in the reduced physical interval
$[0 , \pi]$ of the string.
We therefore propose the following anti-brackets in the fermionic sector:
\begin{equation}
\{\psi^{\mu}_{A}(\sigma), \psi^{\nu}_{B}(\sigma^{\prime})\}
=-i\eta^{\mu\nu}\Lambda_{AB}(\sigma, \sigma^{\prime})\,,
\label{45}
\end{equation}
instead of (\ref{5}) which, when written down explicitly in terms of components, reads
\begin{eqnarray}
\{\psi^{\mu}_{+}(\sigma), \psi^{\nu}_{+}(\sigma^{\prime})\}
&=&\{\psi^{\mu}_{-}(\sigma), \psi^{\nu}_{-}(\sigma^{\prime})\}
=-i\eta^{\mu\nu}\delta_{P}(\sigma - \sigma^{\prime})\,,\nonumber\\
\{\psi^{\mu}_{-}(\sigma), \psi^{\nu}_{+}(\sigma^{\prime})\}
&=&-i\eta^{\mu\nu}\delta_{P}(\sigma + \sigma^{\prime})\,.
\label{46}
\end{eqnarray}
We shall now investigate the consisitency of this structure. Firstly, this
structure of the antibracket relations is completely
consistent with the boundary condition (\ref{7}). To see this explicitly,
we compute the anticommutator of $\psi_{+}(\sigma^{\prime})$ with 
(\ref{7}), the left hand side of which gives
\begin{eqnarray}
-i\left(\delta_{P}(\sigma - \sigma^{\prime})-
\delta_{P}(\sigma + \sigma^{\prime})\right)|_{\sigma = 0, \pi}=
- i \Delta_{-}\left(\sigma , \sigma^{\prime})\right)|_{\sigma = 0, \pi}
= \frac{1}{\pi}\sum_{n\neq 0}
\mathrm{sin}(n\sigma^{\prime})\mathrm{sin}(n\sigma)|_{\sigma = 0, \pi}=0\,,
\label{47}
\end{eqnarray}
where the form of the periodic delta function has been used. 
Not only that, as a bonus, we reproduce the modified form of (\ref{8}).
Observe the occurence of $\delta_{P}(\sigma + \sigma^{\prime})$ rather
than $\delta_{P}(\sigma - \sigma^{\prime})$ in the mixed bracket
$\{\psi_+ , \psi_-\}$, which plays a crucial role in obtaining the
following involutive algebra in the fermionic sector.
Indeed, using (\ref{45}), one can show that 
\begin{eqnarray}
\{\lambda_1(\sigma) , \lambda_1(\sigma^{\prime})\} &=& 4 \left(
\lambda_2(\sigma)\partial_{\sigma}\Delta_{+}
\left(\sigma , \sigma^{\prime}\right) + \lambda_2(\sigma^{\prime})
\partial_{\sigma}\Delta_{-}\left(\sigma , \sigma^{\prime}\right)\right)\,,
\nonumber \\ 
\{\lambda_2(\sigma) , \lambda_2(\sigma^{\prime})\} &=&  \left(
\lambda_2(\sigma^{\prime})\partial_{\sigma}\Delta_{+}
\left(\sigma , \sigma^{\prime}\right) + \lambda_2(\sigma)
\partial_{\sigma}\Delta_{-}\left(\sigma , \sigma^{\prime}\right)\right) \,,
\nonumber \\ 
\{\lambda_2(\sigma) , \lambda_1(\sigma^{\prime})\} &=&  \left(
\lambda_1(\sigma) + \lambda_1(\sigma^{\prime})\right)
\partial_{\sigma}\Delta_{+}\left(\sigma , \sigma^{\prime}\right)
\label{49}
\end{eqnarray}
hold for the fermionic sector.

In order to write down the complete algebra of the super Virasoro
constraints $\chi_{1}(\sigma)$ and $\chi_{2}(\sigma)$, one must take
into account the algebra of constraints between the Bosonic variables.
Interestingly, these have exactly the same structure as the fermionic algebra
(\ref{49})
with the $\lambda$'s being replaced by $\Phi$'s
\footnote{Note that there were some errors in \cite{rb}.}, so that
the complete algebra  of the super Virasoro
constraints also have identical structures :
\begin{eqnarray}
\{\chi_1(\sigma) , \chi_1(\sigma^{\prime})\} &=& 4 \left(
\chi_2(\sigma)\partial_{\sigma}\Delta_{+}
\left(\sigma , \sigma^{\prime}\right) + \chi_2(\sigma^{\prime})
\partial_{\sigma}\Delta_{-}\left(\sigma , \sigma^{\prime}\right)\right)\,,
\nonumber \\ 
\{\chi_2(\sigma) , \chi_2(\sigma^{\prime})\} &=&  \left(
\chi_2(\sigma^{\prime})\partial_{\sigma}\Delta_{+}
\left(\sigma , \sigma^{\prime}\right) + \chi_2(\sigma)
\partial_{\sigma}\Delta_{-}\left(\sigma , \sigma^{\prime}\right)\right) \,,
\nonumber \\ 
\{\chi_2(\sigma) , \chi_1(\sigma^{\prime})\} &=&  \left(
\chi_1(\sigma) + \chi_1(\sigma^{\prime})\right)
\partial_{\sigma}\Delta_{+}\left(\sigma , \sigma^{\prime}\right)\,.
\label{49q}
\end{eqnarray}

\noindent The algebra between the constraints (\ref{280}) now gets
modified to
\begin{eqnarray}
\{\tilde{J}_{1}(\sigma) , \tilde{J}_{1}(\sigma^{\prime})\}
&=&-i(\chi_{1}(\sigma)-2\chi_{2}(\sigma))
\delta_P(\sigma-\sigma^{\prime})\,,\nonumber\\
\{\tilde{J}_{2}(\sigma) , \tilde{J}_{2}(\sigma^{\prime})\}
&=&-i(\chi_{1}(\sigma)+2\chi_{2}(\sigma))
\delta_P(\sigma-\sigma^{\prime})\,,\nonumber\\
\{\tilde{J}_{1}(\sigma) , \tilde{J}_{2}(\sigma^{\prime})\}&=&
-i(\chi_{1}(\sigma)-2\chi_{2}(\sigma))\delta_P(\sigma + \sigma^{\prime})\,.
\label{82a}
\end{eqnarray}
The algebra between $\tilde J(\sigma)$ and $\chi(\sigma)$ can now be computed
 by using the modified bracket (\ref{00}) to get
\begin{eqnarray}
\{\chi_{1}(\sigma) , \tilde{J}_{1}(\sigma^{\prime})\}
&=& - \left(2 \tilde{J}_{1}(\sigma) + \tilde{J}_{1}(\sigma^{\prime})\right)
\partial_{\sigma}\delta_P(\sigma-\sigma^{\prime}) +
\left(2 \tilde{J}_{2}(\sigma) + \tilde{J}_{1}(\sigma^{\prime})\right)
\partial_{\sigma}\delta_P(\sigma + \sigma^{\prime})\,,
\nonumber\\
\{\chi_{1}(\sigma) , \tilde{J}_{2}(\sigma^{\prime})\}
&=& \left(2 \tilde{J}_{2}(\sigma) + \tilde{J}_{2}(\sigma^{\prime})\right)
\partial_{\sigma}\delta_P(\sigma-\sigma^{\prime}) - 
\left(2 \tilde{J}_{1}(\sigma) + \tilde{J}_{2}(\sigma^{\prime})\right)
\partial_{\sigma}\delta_P(\sigma + \sigma^{\prime})\,,
\nonumber\\
\{\chi_{2}(\sigma) , \tilde{J}_{1}(\sigma^{\prime})\}
&=& \left(\tilde{J}_{1}(\sigma) + \frac{1}{2}\tilde{J}_{1}(\sigma^{\prime})\right)\partial_{\sigma}\delta_P(\sigma-\sigma^{\prime}) +
\left(\tilde{J}_{2}(\sigma) + \frac{1}{2}\tilde{J}_{1}(\sigma^{\prime})\right)\partial_{\sigma}\delta_P(\sigma + \sigma^{\prime})\,,
\nonumber\\
\{\chi_{2}(\sigma) , \tilde{J}_{2}(\sigma^{\prime})\}
&=& \left(\tilde{J}_{2}(\sigma) + \frac{1}{2}\tilde{J}_{2}(\sigma^{\prime})\right)\partial_{\sigma}\delta_P(\sigma-\sigma^{\prime}) +
\left(\tilde{J}_{1}(\sigma) + \frac{1}{2}\tilde{J}_{2}(\sigma^{\prime})\right)\partial_{\sigma}\delta_P(\sigma + \sigma^{\prime})\,,\\
\label{82b}
\end{eqnarray}
which clearly displays a new structure for the super-Virasoro algebra.

As a matter of consistency, we write down the hamiltonian of the 
superstring and then study the time evolution of the
$\psi_{\pm}$ modes. This follows easily from the Virasoro functional
$L[f]$ (\ref{33}) by setting $f(\sigma)=e^{im\sigma}$, which gives
\begin{eqnarray}
L_{m} = \frac{1}{4}\int_{-\pi}^{\pi} d\sigma e^{-im\sigma}[\{\Pi(\sigma)+
\partial_{\sigma}X(\sigma)\}^2+2i\psi^{\mu}_{+}\partial_{\sigma}
\psi_{\mu+}]\,.
\label{51}
\end{eqnarray}
Setting $m=0$, gives the hamiltonian
\begin{eqnarray}
H=L_{0} &=& \frac{1}{4}\int_{-\pi}^{\pi} d\sigma [\{\Pi(\sigma)+
\partial_{\sigma}X(\sigma)\}^2+2i\psi^{\mu}_{+}\partial_{\sigma}
\psi_{\mu+}]\nonumber\\
&=& \frac{1}{2}\int_{0}^{\pi} d\sigma[\Pi^2(\sigma)+
\partial_{\sigma}X(\sigma)^2+i(\psi^{\mu}_{+}(\sigma)\partial_{\sigma}
\psi_{\mu+}(\sigma)-\psi^{\mu}_{-}(\sigma)\partial_{\sigma}\psi_{\mu-}
(\sigma))]\,.
\label{52}
\end{eqnarray}
This immediately leads to
\begin{eqnarray}
\dot{\psi_{-}}(\sigma)=\{\psi_{-}(\sigma), H\}
=-\partial_{\sigma}\psi_{-}(\sigma)\quad;\quad
\dot{\psi_{+}}(\sigma)=\{\psi_{+}(\sigma), H\}
=\partial_{\sigma}\psi_{+}(\sigma)\,,
\label{52a}
\end{eqnarray}
which are precisely the equations of motion for the fermionic fields.
One can therefore regard (\ref{00}) and (\ref{46}) as the final 
symplectic structure of the free superstring theory.

%%%%%%%%%%%%%%%%%%%%%%%%%%%%%%%%%%%%%%%%%%%%%%%%%%%%%%%%%%%%%%%%%%%%%%%%%%%%%
%%%%%%%%%%%%%%%%%      The interacting theory
%%%%%%%%%%%%%%%%%%%%%%%%%%%%%%%%%%%%%%%%%%%%%%%%%%%%%%%%%%%%%%
\section{The interacting theory :}
The action for a super string moving in the presence of 
a constant background Neveu-Schwarz two form field 
${\cal F}_{\mu \nu}$ is given by,
\begin{eqnarray}
\label{2}
S &=&
-{ 1 \over 2}
\int_{\Sigma} d^2\sigma \Big(
\eta_{\mu \nu } \partial_a X^\mu \partial^a X^\nu 
\,+ \, \epsilon^{ab}{\cal F}_{\mu \nu} \partial_a X^\mu \partial_b X^\nu  \nonumber\\
&-& i {\overline \psi}^\mu \rho^a \partial_ a \psi_\mu 
+ i {\cal F}_{\mu\nu} {\overline \psi}^\mu \rho_b \epsilon^{ab}  \partial_ a \psi^\nu 
 \Big)\,.
\label{53}
\end{eqnarray}
The bosonic and fermionic sectors decouple. We 
consider just the fermionic sector since the bosonic sector 
was already discussed \cite{rb}. In component the fermionic sector reads
\begin{equation}
S_F \,=\,
{i \over 2}
\int_{\Sigma} d\tau d\sigma \Big(
\psi^\mu_{-}  \partial_+\, \psi_{-\,\mu }\,+\,
\psi^\mu_{+}  \partial_- \, \psi_{+\,\mu}\,
\,-\, {\cal F}_{\mu\nu} \psi^\mu_{-}  \partial_+ \, \psi^\nu_{-}
+ {\cal F}_{\mu\nu} \psi^\mu_{+}  \partial_- \, \psi^\nu_{+} \,\Big)\,.
\label{54}
\end{equation}
The minimum action principle $\delta S = 0$ leads to a volume term
that vanishes when the equations of motion hold, and also to a surface term
\begin{equation}
\label{a3}
\Big( \psi^\mu_{-} (\eta_{\mu\nu} - {\cal F}_{\mu\nu}) \delta \psi^\nu_{-}\,-\,
\psi^\mu_{+} ( \eta_{\mu\nu} +  {\cal F}_{\mu\nu}) \delta \psi^\nu_{+}
\Big)\vert_{0}^\pi\,\,=\,\,0\,\,.
\end{equation}
It is not possible to find non trivial boundary conditions
involving $\psi^{\mu}_{-}$ and $\psi^{\mu}_{+}$ that makes the above surface
term vanish.  However, the addition of a boundary term
\cite{haggi}, \cite{NL}
\begin{eqnarray}
S_{bound}=\frac{i}{2\pi\alpha^{\prime}}\int_{\Sigma}d\tau d\sigma
 \left(\cal F_{\mu\nu}\psi^{\mu}_{+}\partial_{-}\psi^{\nu}_{+}\right) 
\label{550}
\end{eqnarray}
makes it possible to find a solution to the boundary condition.
Addition of this term to $S_{F}$ leads to the total action:
\begin{eqnarray}
S=\frac{-i}{4\pi\alpha^{\prime}}\int_{\Sigma}d\tau d\sigma
 \left( \psi^{\mu}_{-}E_{\nu\mu}\partial_{+}\psi^{\nu}_{-}+
\psi^{\mu}_{+}E_{\nu\mu}\partial_{-}\psi^{\nu}_{+}\right)\,, 
\label{551}
\end{eqnarray}
where $E^{\mu\nu} \,=\, \eta^{\mu\nu} \, 
+ {\cal F}^{\mu\nu}$.
The corresponding boundary term coming from $\delta S=0$ is given by
\begin{eqnarray}
\left(\psi^{\mu}_{-}E_{\nu\mu}\delta\psi^{\nu}_{-}-
\psi^{\mu}_{+}E_{\nu\mu}\delta\psi^{\nu}_{+}\right)|_{0}^{\pi}=0.
\label{552}
\end{eqnarray}

The above condition is satisfied by the following conditions that
preserve supersymmetry \cite{pi} at the string endpoints
 $\sigma=0$ and $\sigma=\pi$:
\begin{eqnarray}
 E_{\nu\mu}\,  \psi^\nu_{+} (0,\tau)\,&=&\,
 E_{\mu\nu} \, \psi^\nu_{-} (0,\tau) \,,\nonumber \\
 E_{\nu\mu}\,  \psi^\nu_{+} (\pi,\tau ) \,&=&\,\lambda
 E_{\mu\nu} \, \psi^\nu_{-} (\pi, \tau )\,\,,
\label{55}
\end{eqnarray}
 where $\lambda = \pm 1 \,$ with 
the plus sign corresponding to Ramond
boundary condition and the minus corresponding to the Neveu-Schwarz case.
Here too we work with Ramond boundary conditions.
Now the BCs are recast as
\begin{eqnarray}
 \left(E_{\nu\mu}\,  \psi^\nu_{(+)} (\sigma,\tau)\, 
-\, E_{\mu\nu} \, \psi^\nu_{(-)} (\sigma,\tau)\right)
\vert_{\sigma = 0, \pi} \, = \, 0\,.
\label{56}
\end{eqnarray}
This nontrivial BC leads to a modification in the original
(naive) (\ref{5}) DBs. The $\{\psi^\mu_{(+)} (\sigma,\tau) , 
\psi^\nu_{(+)} (\sigma^{\prime},\tau)\}_{DB}$ is the same as 
that of the free string (\ref{5}). We therefore make an ansatz
\begin{eqnarray}
 \{\psi^\mu_{+} (\sigma,\tau) , 
\psi^\nu_{-} (\sigma^{\prime},\tau)\}_{DB}\, = \, 
C^{\mu\nu}\delta_{P}\left(\sigma + \sigma^{\prime}\right)\,.
\label{57}
\end{eqnarray}

Taking brackets between the BCs (\ref{56}) and 
$\psi^\gamma_{-} (\sigma^{\prime})$  we get
\begin{eqnarray}
 E_{\nu\mu}\,  C^{\nu \gamma}
\, = \, -i \, E_{\mu\gamma}.
\label{58}
\end{eqnarray}
Solving this, we find
\begin{eqnarray}
 C^{\mu \nu}\, = \, -i \, 
\left[\left(1 - {\cal F}^2\right)^{-1}\right]^{\mu \rho}\, 
 E_{\rho \gamma}\, E^{\gamma \nu}.
\label{58a}
\end{eqnarray}
One can also take brackets between the BCs (\ref{56}) and 
$\psi^\gamma_{+} (\sigma^{\prime})$, which yields
\begin{eqnarray}
 C^{\nu \mu}\, = \, -i \, 
\left[\left(1 - {\cal F}^2\right)^{-1}\right]^{\mu \rho}\, 
 E_{\gamma \rho}\, E^{\nu \gamma}.
\label{58b}
\end{eqnarray}
Although the expressions (\ref{58a}) and (\ref{58b}) look
different, they are actually the same as can be easily verified.
Finally we can write the matrix 
$C = \{C^{\mu \nu}\}$ more compactly as 
\begin{eqnarray}
 C \, = \, -i \, 
\left[\left(1 - {\cal F}^2\right)^{-1}\left(1 + {\cal F}\right)^{2}\right].
\label{58y}
\end{eqnarray}

We therefore get the following modification:
\begin{eqnarray}
\{\psi^\mu_{+} (\sigma,\tau) , 
\psi^\nu_{-} (\sigma^{\prime},\tau)\}_{DB}\, = \, 
-i\, \left[\left(1 - {\cal F}^2\right)^{-1}\right]^{\mu \rho}
\,  E_{\rho \gamma}\, E^{\gamma \nu}
\delta_{P}\left(\sigma + \sigma^{\prime}\right)\,,
\label{60}
\end{eqnarray}
which also reduces to those of \cite{godinho}, upto the 
$\delta_{P}\left(\sigma + \sigma^{\prime}\right)$ factor.
Finally, note that in the limit ${\cal F}_{\mu \nu} \rightarrow 0$ (\ref{60}), the last of  (\ref{46}) is reproduced.

%%%%%%%%%%%%%%%%%%%%%%%%%%%%%%%%%%%%%%%%%%%%%%%%%%%%%%%%%%
%%%%%%%%%%%%% Conclusions %%%%%%%%%%%%%%%%%%%%%%%%%%%%%%%%
\section{Conclusions}
In this paper, we have derived the expressions for a non(anti)
commutative algebra for an open superstring. The interesting
thing to note that, unlike the bosonic case, we get
an anticommutative structure in the fermionic sector even for 
the free superstring. Our results differ from those in \cite{godinho}
and are mathematically consistent which is reflected from the 
closure of the constraint algebras. The analysis of this paper is a
direct generalisation of \cite{rb}, where only bosonic string was considered.

The origin of any modification in the usual canonical algebra
is the presence of boundary conditions. This phenomenon is quite well
known for a free scalar field subjected to periodic boundary conditions.
Besides this method was also used earlier by \cite{hrt} in the context of 
Nambu-Goto formulation of the bosonic string.
We show that the same also holds true in the fermionic sector of 
the conformal gauge fixed free superstring, where the boundary conditions
become periodic once we extend the domain of definition of the length
of the string from $[0, \pi]$ to $[-\pi, \pi]$.  This mathematical
trick leads to a modification where the usual Dirac delta function
gets replaced by a periodic delta function. Eventually one constructs
the appropriate ``delta function" for the physical interval $[0, \pi]$
of the string to write down the basic symplectic structure. Interestingly,
here we get a $2 \times 2$ matrix valued ``delta function" appropriate
for the two component Majorana spinor which is in contrast to the
bosonic case, where one has a single component ``delta function"
$\Delta_{+}(\sigma , \sigma^{\prime})$ satisfying Neumann boundary condition
\cite{hrt} , \cite{rb}.
This symplectic structure, interestingly, leads to a new involutive
structure for the super-Virasoro algebra at the classical level.
The corresponding quantum version and its implications are being investigated.

The same technique is adopted for the interacting case also.
Here the boundary condition is more involved and leads to
 a more general type of non(anti)-commutativity that has been observed
before. However, our results are once again different from 
the existing results since we get a  periodic delta function
instead of the usual delta function, apart from the relative sign
of $\sigma , \sigma^{\prime}$. This change of relative sign indeed plays a 
crucial role in the internal consistency of our analysis.
Further, the interacting results go over smoothly to the free
case once the interaction is switched off.

%%%%%%%%%%%%%%%%%%%%%%%%%%%%%%%%%%%%%%%%%%%%%
%%%%%%%%%% acknowledge
%%%%%%%%%%%%%%%%%%%%%%%%%%
\section*{Acknowledgements }

This work was supported by a grant under the Indo-South African research
agreement between the department of Science and Technology, Government
of India and the South African National Research Foundation. BC would like
to thank the Institute of Theoretical Physics, Stellenbosch University
for their hospitality during the period when part of this work was 
completed. FGS would like to thank the S.N.Bose National Centre For
Basic Sciences, for their hospitality in the periods that parts of this 
work were completed. SG and AGH would like to thank Prof. A. P. 
Balachandran for some useful discussions. The authors would also like
to thank the referee for useful comments.

%\newpage


\begin{thebibliography}{99}
\bibitem{sw}N. Seiberg and E. Witten, JHEP 09 032(1999).
\bibitem{dn}For a review, see M.R.Douglas and N.A.Nekrasov,
arXiv: hep-th/0108158; R.J. Szabo,  arXiv: [hep-th/0109162].
\bibitem{green}M.B. Green, J.H. Schwarz, E. Witten,
{\it Superstring theory Vol.1}, Cambridge University Press, 1987 ;
D. L\"{u}st, S. Theisen, {\it Lectures on String Theory},
Springer-Verlag 1989.
\bibitem{chu}C.-S.Chu and P.-M.Ho, Nucl.Phys. B550 (1999) 151.
\bibitem{som}V. Schomerus, JHEP 06 (1999) 030.
\bibitem{di}P.A.M. Dirac, {\it Lectures on Quantum Mechanics} (Yeshiva
University Press, New York, 1964).
\bibitem{rb}R. Banerjee, B. Chakraborty, S. Ghosh,
Phys.Lett. B537 (2002) 340-350; [hep-th/0203199],\\
R. Banerjee, B. Chakraborty, K. Kumar Nucl.Phys. B668 (2003)179;
 [hep-th/0306122].
\bibitem{godinho}N.R.F. Braga and C.F.L. Godinho,
 Phys.Lett. B570 (2003) 111-117; [hep-th/0306163].
\bibitem{ard}F. Ardalan, H. Arfaei and M.M. Sheikh-Jabbari, JHEP 9902 
(1999) 016; W.T. Kim and J.J. Oh, 
Mod.Phys.Lett. A15 (2000) 1597; C.-S. Chu and P.-M. Ho, 
Nucl.Phys. B568 (2000) 447.
\bibitem{bra}N.R.F. Braga and C.F.L. Godinho, [hep-th/0110297].
\bibitem{tez}M.M. Sheikh-Jabbari and A.Shirzad, Eur. Phys. Jour. C19 (2001) 383;
K-I. Tezuka, [hep-th/0201171].
\bibitem{hrt}A.J. Hanson, T. Regge and C. Teitelboim, {\it Constrained
Hamiltonian System}, Roma, Accademia Nazionale Dei  Lincei, (1976).
\bibitem{zab}M. Zabzine, JHEP 0010 (2000) 042.
\bibitem{and}M.A De Andrade, M.A. Santos and I.V.Vancea, JHEP 0106 (2001) 026.
%\bibitem{holt}See for example J.W.van Holten,
% {\it Aspects of BRST Quantisation },
%hep-th/0201124.
%\bibitem{mk}M.Kaku, {\it Introduction to Superstring Theory}, Springer-Verlag, 1988.
%\bibitem{pol}J. Polchinski, {\it String Theory}, Vol. I, Cambridge University Press, 1998.
\bibitem{haggi}P. Haggi-Mani, U. Lindstrom, M. Zabzine,
 Phys.Lett. B483 (2000) 443-450.
\bibitem{NL}U. Lindstrom, M. Rocek, P. van Nieuwenhuizen,
"Consistent Boundary Conditions for open strings", hep-th/0211266.
\bibitem{pi}C.-S.Chu and P.-M.Ho, Nucl.Phys. B568 (2000) 447.
\bibitem{fad}L. Faddeev, R. Jackiw, Phys.Rev.Lett. 60 (1988) 1692.
\bibitem{schwinger}J. Schwinger, Lester L. DeRaad,Jr, Kimball A. Milton,
Wu-yang Tsai, {\it Classical Electrodynamics}, Advanced Book Program,
Perseus Books.
\end{thebibliography}
\end{document}